\documentstyle[12pt,psfig,rotating]{article}
\newlength{\dinmargin}
\newlength{\dinwidth}
\mathchardef\less=316
\mathchardef\greater=318
\def\gev{{\rm Ge}\kern-1.pt{\rm V}}
\def\mev{{\rm Me}\kern-1.pt{\rm V}}

\def\econe{E_{\rm cone}}

\def\obs{{\rm obs}}
\def\DA{{\scriptscriptstyle {\rm DA}}}
\def\xDA{{x_\DA}}
\def\xDAs{{x_\DA^*}}
\def\yDA{{y_\DA}}
\def\QsqDA{{Q^2_\DA}}
\def\QsqDAs{{Q^{2*}_\DA}}
\def\gamraw{{\gamma_{\rm raw}}}
\def\thetae{{\theta_e}}
\def\RMS{{r.\,m.\,s.\ }}

\newbox\struttbox
\setbox\struttbox=\hbox{\vrule height1.65ex depth.485ex width0pt}
\def\strutt{\relax\ifmmode\copy\struttbox\else\unhcopy\struttbox\fi}
\def\stru#1#2{\relax\ifmmode\hbox{\vrule height#1 depth#2 width0pt}
\else\vrule height#1 depth#2 width0pt\fi}
\def\uline#1{$\underline{\hbox{#1\strutt}}$}
\mathchardef\eql=61
\def\cha{{\phantom{0}}}

\catcode`\@=11
\def\gappr{\mathpalette\under@rel{>\approx}}
\def\lappr{\mathpalette\under@rel{<\approx}}
\def\gsim{\mathpalette\under@rel{>\sim}}
\def\lsim{\mathpalette\under@rel{<\sim}}
\def\under@rel#1#2{\under@@rel#1#2}
\def\under@@rel#1#2#3{\mathrel{\mathop{#1#2}\limits_{#1#3}}}

\def\parenbar{\mathpalette\p@renb@r}
\def\p@renb@r#1#2{\vbox{%
  \ifx#1\scriptscriptstyle \dimen@.7em\dimen@ii.2em\else
  \ifx#1\scriptstyle \dimen@.8em\dimen@ii.25em\else
  \dimen@1em\dimen@ii.4em\fi\fi \offinterlineskip
  \ialign{\hfill##\hfill\cr
    \vbox{\hrule width\dimen@ii}\cr
    \noalign{\vskip-.3ex}%
    \hbox to\dimen@{$\mathchar300\hfil\mathchar301$}\cr
    \noalign{\vskip-.3ex}%
    $#1#2$\cr}}}
\catcode`\@=12 

\begin{document}
\vspace{1 cm}
\begin{titlepage}
\title{\bf
\stru{2.3ex}{2.3ex} Comparison of ZEUS Data \\
\stru{2.3ex}{2.3ex} with Standard Model Predictions \\
\stru{2.3ex}{2.3ex} for $e^+p \rightarrow e^+X$  Scattering at High $x$ and $Q^2$}
\vskip2.cm
\author{ZEUS Collaboration}
\date{}
\maketitle

\vspace{5 cm}

\begin{abstract}
\centerline{\vbox{\hsize13.7cm
\noindent
Using the ZEUS detector at HERA, we have studied the reaction
$e^+ p\rightarrow e^+X$ for $Q^2>5000\,\gev^2$ with a
$20.1\,{\rm pb}^{-1}$ data sample collected during the
years 1994 to 1996.
For $Q^2$ below 15000\,GeV$^2$, the data are in good
agreement with Standard Model expectations.
For $Q^2 > 35000\;\gev^2$, two events are observed while
$0.145 \pm 0.013$ events are expected.
A statistical analysis of a large ensemble of simulated
Standard Model experiments indicates that with probability 6.0\%,
an excess at least as unlikely as that observed
would occur above {\it some} $Q^2$ cut.
For $x>0.55$ and $y>0.25$, four events are observed where
$0.91\pm 0.08$ events are expected.
A statistical analysis of
the two-dimensional distribution of the events in $x$ and $y$
yields a probability of 0.72\% for the region $x>0.55$ and $y>0.25$
and a probability of 7.8\% for the entire $Q^2>5000\,$GeV$^2$ data
sample.
The observed excess above Standard Model expectations
is particularly interesting
because it occurs in a previously unexplored kinematic region. }}
\end{abstract}
\begin{center}
To be published in Zeitschrift f\"ur Physik C.
\end{center}
\vskip -23cm
\centerline{{\tt DESY 97-025}\hfill{\tt ISSN 0418-9833}}
\setcounter{page}{0}
\thispagestyle{empty}   
\end{titlepage}
\begin{center}
{                      \Large  The ZEUS Collaboration              }
\end{center}
  J.~Breitweg,
  M.~Derrick,
  D.~Krakauer,
  S.~Magill,
  D.~Mikunas,
  B.~Musgrave,
  J.~Repond,
  R.~Stanek,
  R.L.~Talaga,
  R.~Yoshida,
  H.~Zhang  \\
 {\it Argonne National Laboratory, Argonne, IL, USA}~$^{p}$
% ...
\newpage
% ...
$^{\    2}$ now at Univ. of Salerno and INFN Napoli, Italy \\
% ...
$^{  21}$ now at ATLAS Collaboration, Univ. of Munich \\
% ...
                                                           %
                                                           %
\newpage   
                                                           %
                                                           %
\begin{tabular}[h]{rp{14cm}}
$^{a}$ &  supported by the Natural Sciences and Engineering Research
          Council of Canada (NSERC)  \\
$^{b}$ &  supported by the FCAR of Qu\'ebec, Canada  \\
$^{c}$ &  supported by the German Federal Ministry for Education and
          Science, Research and Technology (BMBF), under contract
          numbers 057BN19P, 057FR19P, 057HH19P, 057HH29P, 057SI75I \\
$^{d}$ &  supported by the MINERVA Gesellschaft f\"ur Forschung GmbH,
          the German Israeli Foundation, and the U.S.-Israel Binational
          Science Foundation \\
$^{e}$ &  supported by the German Israeli Foundation, and
          by the Israel Science Foundation
  \\
$^{f}$ &  supported by the Italian National Institute for Nuclear Physics
          (INFN) \\
$^{g}$ &  supported by the Japanese Ministry of Education, Science and
          Culture (the Monbusho) and its grants for Scientific Research \\
$^{h}$ &  supported by the Korean Ministry of Education and Korea Science
          and Engineering Foundation  \\
$^{i}$ &  supported by the Netherlands Foundation for Research on
          Matter (FOM) \\
$^{j}$ &  supported by the Polish State Committee for Scientific
          Research, grant No.~115/E-343/SPUB/P03/120/96  \\
$^{k}$ &  supported by the Polish State Committee for Scientific
          Research (grant No. 2 P03B 083 08) and Foundation for
          Polish-German Collaboration  \\
$^{l}$ &  partially supported by the German Federal Ministry for
          Education and Science, Research and Technology (BMBF)  \\
$^{m}$ &  supported by the German Federal Ministry for Education and
          Science, Research and Technology (BMBF), and the Fund of
          Fundamental Research of Russian Ministry of Science and
          Education and by INTAS-Grant No. 93-63 \\
$^{n}$ &  supported by the Spanish Ministry of Education
          and Science through funds provided by CICYT \\
$^{o}$ &  supported by the Particle Physics and
          Astronomy Research Council \\
$^{p}$ &  supported by the US Department of Energy \\
$^{q}$ &  supported by the US National Science Foundation \\
\end{tabular}
                                                           %
\newpage

\section{Introduction}

\label{sec-intro}

Deep--inelastic scattering (DIS) of leptons on nucleons has been an
important tool for understanding nucleon structure and many elements of
the Standard Model, including both the electroweak interaction and quantum
chromodynamics (QCD). At the HERA collider, DIS processes are
being studied at a center of mass energy $\sqrt{s}=300\,$GeV
and at $Q^2$ (the negative of the square of the four-momentum transfer)
exceeding the squares of the weak vector boson masses. In this regime,
lepton--nucleon scattering allows unique and sensitive tests of
the Standard Model as well as of certain extensions to it~\cite{SM_ext}.

This paper presents results from $e^+p$ running with the ZEUS detector
during the years 1994 to 1996, at proton and positron beam energies of
$E_p=820\,\gev$ and $E_e=27.5\,\gev$.
With the integrated luminosity of $20.1\,{\rm pb}^{-1}$
collected in this period, it has become
possible to study the reaction $e^+p\rightarrow e^+X$ in the region
where the expected DIS cross section is in the subpicobarn range.
This region of high $Q^2$ and $x$ (the Bjorken scaling
variable) has never before been explored.
The above reaction is understood to be a positron--quark collision with
center--of--mass energy $\sqrt{xs}$. Initial cross section measurements
by the ZEUS~\cite{ref-ZEUS-hiq293}
and H1~\cite{ref-H1-hiq294} collaborations are in good agreement with
Standard Model expectations for $Q^2$ up to about $10^4\,$GeV$^2$.
In this paper, we report on a more sensitive search for
deviations from Standard Model predictions in the region $Q^2>5000\,$GeV$^2$.


\section{Neutral Current Deep--Inelastic Scattering}

\label{sec-ncdis}

The reaction studied is:
\begin{equation}
e^{+}+p\rightarrow e^{+}+X  \label{eq-react}
\end{equation}
where $X$ represents the final state hadronic system. In the high $Q^2$
regime, the Standard Model neutral current (NC)
cross section for~(\ref{eq-react}) depends on well--measured
electroweak parameters and on the parton densities in the proton. Though the
latter have not yet been measured at high $Q^2$, perturbative Quantum
Chromodynamics (pQCD) predicts
their values through evolution from high--precision measurements made at
lower $Q^2$ values.

The Born cross section \cite{csectDIS} for the NC DIS reaction (\ref
{eq-react}) with unpolarized beams is\footnote{%
We neglect the contribution to the cross section (\ref{eq-csect}) of the
longitudinal structure function, ${\cal F}_L$, which we estimate from
pQCD and the parton densities\cite{flpred} to be less than $1\%$
in the kinematic range under study.}
\begin{equation}
\frac{d^2\sigma }{dx\ dQ^2}=\frac{2\pi \alpha ^2}{xQ^4}\left\{ Y_{+}(y)\,%
{\cal F}_2(x,Q^2)-Y_{-}(y)\,x{\cal F}_3(x,Q^2)\right\} \;,  \label{eq-csect}
\end{equation}
where $\alpha $ is the electromagnetic coupling. The cross section is given
in terms of $Q^2$ and the DIS scaling variables $x$ and $y=Q^2/sx$. In the
region of large $x$ and $Q^2$ studied here, the parity--violating $x{\cal F}_3$
term substantially reduces the $e^{+}p$ cross section, while
increasing the cross section for $e^{-}p$ scattering (where the second term
has positive sign).  The explicit $y$--dependence, which is due to the
helicity dependence of electroweak interactions, is contained in the
functions
\begin{equation}
Y_{\pm }(y)=1\pm (1-y)^2\;,  \label{ydep}
\end{equation}
while the dependence on the quark structure of the proton, and on the $Z^0$
propagator is absorbed in the (positive) structure functions:
\begin{equation}
\left(
\begin{array}{c}
{\cal F}_2(x,Q^2) \\
x{\cal F}_3(x,Q^2)
\end{array}
\right) =x\sum_{q={\rm quarks}}\left(
\begin{array}{c}
C_2^q(Q^2)[q(x,Q^2)+\overline{q}(x,Q^2)] \\
C_3^q(Q^2)[q(x,Q^2)-\overline{q}(x,Q^2)]
\end{array}
\right)  \label{eq-stf}
\end{equation}
written in terms of the quark densities in the proton
($q=u,\;d,\;c,\;s,\;t,\;b$)
and the corresponding antiquark densities $\overline{q}$.
For $e^{+}p$ scattering, the $Q^2$--dependent coefficient
functions, $C_2^q$ and $C_3^q$, are given by:
\begin{equation}
\begin{array}{rrl}
C_2^q(Q^2)=&e_q^2 & -2e_qv_qv_e\chi_Z+(v_q^2+a_q^2)(v_e^2+a_e^2)\chi _Z^2 \\
C_3^q(Q^2)=&      & -2e_qa_qa_e\chi_Z+(2v_qa_q)(2v_ea_e)\chi_Z^2
\end{array}\label{eq-coefun}
\end{equation}
with
\begin{equation}
\chi _Z=\frac 1{4\sin ^2\theta _w\cos ^2\theta _w}\frac{Q^2}{Q^2+M_Z^2}\;.
\label{eq-chiZ}
\end{equation}
In eqs. \ref{eq-coefun} and \ref{eq-chiZ}, $M_Z$ is the $Z^0$ mass, $e_q$
is the quark charge in units of the positron charge, $v_q=(T_{3q}-2e_q\sin
^2\theta _w)$ and $a_q=T_{3q}$ are the vector and axial vector couplings of
the quark to the $Z^0$, $v_e$ and $a_e$ are the corresponding electron
couplings, $\theta _w$ is the weak mixing angle, and $T_3$ is the third
component of the weak isospin. All relevant electroweak parameters have been
measured to high precision~\cite{EWpar}.

The QCD--evolved structure functions~\cite{StrF} of equation (\ref{eq-stf}),
evaluated at a given $x$ at high $Q^2$, depend on quark and gluon
densities in the proton measured at lower values of $Q^2$ and higher values
of $x$. At high $x$,  $u$ quarks give the dominant
contribution to the cross section because they have the
largest density~\cite{BEBCdoveru} and because $e_u=2/3$. In addition,
the antiquark ($\overline{q}$) density is small \cite{CCFRqbar}.

Uncertainties in the Born-level $e^+ p$ DIS
cross section predictions in this region of high $%
x $ and $Q^2$ are estimated to be about $6.5\%$
(see Section~\ref{sec-smerr}), mainly due to uncertainties in the
evolved quark densities.

It should be noted that an anomalously high cross section for the
production of jets with high transverse energy in $p\overline p$ collisions,
as recently reported by the CDF collaboration \cite{CDFanom}, can be
explained by adjusting the gluon density in the proton \cite{CTEQhiEt}
(which raises the rate of gluon--quark collisions at high $x$), rather than
by adjusting quark densities. This variation of the gluon density, however,
has only a small effect on the cross section predictions relevant to this
paper (see Section~\ref{sec-smerr}).


\section{ZEUS Detector and Monte--Carlo Simulation}

\label{sec-zeus}


\subsection{Experimental Setup}

\label{ssec-expsetup}

A description of the ZEUS detector can be found in references~\cite
{sigtot_photoprod,Detector}. The primary components used in this analysis
were the compensating uranium--scintillator calorimeter, the central
tracking detector, and the luminosity detector.

The calorimeter~\cite{CAL} is divided into three parts, forward (FCAL)
covering the polar angle\footnote{%
The right-handed ZEUS coordinate system is centered on the nominal
interaction point ($Z=0$) and defined with the $Z$ axis pointing in the
proton beam direction, and the horizontal $X$ axis pointing towards the
center of HERA.} interval $2.6^{\circ }\less\theta \less 37^{\circ }$,
barrel (BCAL: $37^{\circ }\less\theta \less129^{\circ }$) and rear (RCAL: $%
129^{\circ }\less\theta \less176.1^{\circ }$). The calorimeters are
subdivided into towers which each subtend solid angles from $0.006$ to $0.04$
steradians. Each tower is longitudinally segmented into an electromagnetic
(EMC) section and two hadronic (HAC) sections (one in RCAL). Each HAC section
consists of a single cell, while the EMC
section of each tower is further subdivided transversely
into four cells (two in RCAL). In test beam conditions, for particle
energies up to $120\,\gev$,
energy resolutions of $\sigma _E/E\eql 18\%/\sqrt{E(\gev)%
}$ for electrons and $\sigma _E/E\eql35\%/\sqrt{E(\gev)}
$ for hadrons have been measured.
The cell-to-cell variations
in the energy calibration are approximately $2\%$ for the EMC cells
and $3\%$ for HAC cells. The FCAL and BCAL energy scales are presently
understood to an accuracy of $3\%$.
The time resolution is below $1\,{\rm ns}$ for energy
deposits greater than $4.5\,\gev$.
The impact point of the scattered positron
at the calorimeter, determined using pulse height sharing, has a
resolution of about 1~cm.

In the physics analysis, only those calorimeter cells with energy deposits
above thresholds of $60\,\mev$ and $110\,\mev$ for EMC and HAC cells
respectively were used.

The central tracking chamber (CTD) \cite{CTD} operates in a $1.43\,{\rm T}$
solenoidal magnetic field. It is a drift chamber consisting
of 72 cylindrical layers, organized into 9 superlayers. A momentum measurement
requires a track to pass through at least two superlayers, corresponding to
a polar angle region of $15^{\circ }\less\theta \less 164^{\circ }$. The
transverse momentum resolution is $\sigma (p_t)/p_t=\left[ 0.005p_t(\gev)
\right] \oplus 0.016$ for full length tracks. For full length tracks
with momenta $p>5\;\gev$ the vertex resolution is $0.1\,%
{\rm cm}$ in the transverse plane and $0.4\,{\rm cm}$ along $Z$.

Events were filtered online by a three--level trigger system~\cite{Detector}.
The trigger criteria used in this analysis relied primarily on the energies
measured in the calorimeter. The first level trigger decision was based on
electromagnetic energy and total transverse energy ($E_t$). The second level
trigger rejected backgrounds (mostly $p$--gas interactions) for which the
calorimeter timing was inconsistent with an $ep$ interaction. In addition,
the second level trigger applied increased $E_t$ thresholds and also
required a minimum value of $E-p_Z$ (see Section~\ref{sec-evsel}),
where $E$ and $p_Z$ are the summed energy and $Z$-component of the
momentum measured in the calorimeter. The third level trigger applied more
stringent timing cuts as well as increased energy and $E-p_Z$ thresholds.
In all cases, the requirements were less stringent than those imposed by the
offline event selection.

The luminosity was measured by the rate of high energy photons from the
process $ep\to ep\gamma $ detected in a lead--scintillator calorimeter~\cite
{lumi} located at $Z=-107\,{\rm m}$. The uncertainty associated with
luminosity measurements is
addressed in section \ref{sec-smerr}.


\subsection{Monte Carlo Simulation}

\label{ssec-mc}

NC DIS events were simulated using the {\sc meps} option of {\sc lepto} \cite
{lepto} interfaced to {\sc heracles} \cite{heracles} via {\sc django} \cite
{django} and the MRSA parton distribution set \cite{MRSA}. The event
simulation included electroweak radiative corrections, leading order QCD
effects and parton showers. Hadronization was simulated with
{\sc jetset}\cite{pythia}.

Large samples of simulated photoproduction events\cite{ZEUSphoto} were used
for background studies. Samples of both direct and resolved photoproduction
events (including the production of $c\overline{c}$ and $b\overline{b}$ pairs)
were generated
using both {\sc pythia} \cite{pythia} and {\sc herwig} \cite{herwig}. Direct
and resolved photoproduction of events with prompt photons were simulated
with {\sc herwig}. Production of $W$ and $Z$ bosons was studied using the
{\sc epvec} \cite{epvec} generator. Finally, the processes $\gamma \gamma
\rightarrow e^{+}e^{-}$ and $\gamma \gamma \rightarrow \tau ^{+}\tau ^{-}$
were simulated using {\sc zlpair}~\cite{zlpair}.

All MC events were passed through a {\sc geant} \cite{geant} based simulation
of the ZEUS detector and trigger, and analyzed with the same
reconstruction and offline selection procedures as the data.


\section{Positron Identification and Event Kinematics}

\label{sec-kine}

A key signature of high $Q^2$ $e^+p\rightarrow e^+X$
events is an isolated high
transverse momentum positron. In order to identify and reconstruct this
positron, while rejecting events in which other final state particles mimic a
positron, an algorithm was used which combines calorimeter and CTD
information.

In a first step, the calorimeter cells are clustered by
joining each cell to
the highest energy cell among its adjacent neighbours. All clusters are
evaluated as positron candidates. The cluster energy, $E_{{\rm clu}}$, is
the sum of the cell energies belonging to the cluster.  The cluster angle,
$\theta_{{\rm clu}}$, is set equal to the
polar angle obtained by joining the energy-weighted
mean position of the cluster with the event vertex obtained from the tracks
measured with the CTD.  For candidates with polar angle\footnote{%
We do not consider candidates with $\theta_{{\rm clu}}>164^{\circ}$ (which are also
beyond the CTD acceptance limit), since they correspond to $Q^2$ values
below the range of this analysis.} within the CTD acceptance
($\theta_{{\rm clu}} >17.2^{\circ}$%
), a matching track is required. A track is considered to match if the
distance of closest approach (DCA) between the extrapolation of the track
into the calorimeter and the position of the cluster center is less than $%
10\,{\rm cm}$, where the \RMS resolution on the DCA is 1.8 cm.

In the second step, several quantities, $\xi _i$, are calculated for
each positron candidate: the fraction of the
cluster energy in the HAC sections of the calorimeter,
the parameters related to
lateral energy profiles, and the total energy ($\econe$)
in all calorimeter cells not associated with the cluster
but lying within an $\eta,\phi $ (pseudorapidity,azimuth) cone of radius 0.8
centered on the cluster.  If a matching track is present, we
also evaluate the polar and azimuthal angle differences between the track
and the cluster position, and the quantity $1/E_{{\rm clu}}-1/{P_{{\rm trk}}}
$, where ${P_{{\rm trk}}}$ is the track momentum.

Finally, we transform each $\xi _i$ into a quality factor $Q(\xi_i)$.
Candidates are accepted as positrons if the product of the
$Q(\xi_i)$ exceeds a threshold determined from Monte Carlo studies. The
efficiency for finding positrons in a neutral current
DIS\ sample with $Q^2>5000\,\gev^2$
is $91\%$. In accepted events, the positron energy, $E_e^\prime$, is set
equal to the cluster energy, $E_{{\rm clu}}$, and the positron angle,
$\thetae$, is set equal to $\theta_{{\rm clu}}$.
The resolution in $\thetae$ is typically better than $0.3^\circ$.

For each event with an accepted positron, the following global event
quantities were calculated from the energy deposits in the calorimeter:
\begin{eqnarray}
p_t &=&\sqrt{\left( \sum_ip_X^i\right) ^2+\left( \sum_ip_Y^i\right) ^2}\;,
\nonumber \\
E-p_Z &=&\sum_i\left( E^i-p_Z^i\right) \;,  \nonumber \\
E_t &=&\sum_i\sqrt{(p_X^i)^2+(p_Y^i)^2}\;, \\
(p_t)_{{\rm had}} &=&\sqrt{\left( {\sum_i}^\prime p_X^i\right) ^2+\left( {%
\sum_i}^\prime p_Y^i\right) ^2}\;,  \nonumber \\
(E-p_Z)_{{\rm had}} &=&{\sum_i}^\prime \left( E^i-p_Z^i\right) \;,
\nonumber  \label{eq-sumhad}
\end{eqnarray}
where the sums run over all calorimeter cells with energy deposits above
threshold and the $\vec{p\,}^i$ are the momenta assigned to each calorimeter
cell (calculated assuming zero mass with the direction obtained from the
cell center and the measured vertex position). The primed sums
exclude the cells associated with the positron.

To describe the hadronic system, we use the angle, $\gamraw$, and energy,
$E_q $, defined as
\begin{equation}
\cos \gamraw ={\frac{(p_t)_{{\rm had}}^2-(E-p_Z)_{{\rm had}}^2}{(p_t)_{{\rm %
had}}^2+(E-p_Z)_{{\rm had}}^2}}\qquad {\rm and}\qquad E_q={\frac{(p_t)_{{\rm %
had}}}{\sin \gamraw}}\;.  \label{eq-gamdef}
\end{equation}

Resolution effects and systematic shifts of $\gamraw$ have been
studied with MC simulations.  The reconstructed $\gamraw$ is
systematically higher than the generated
value by about $2.7^\circ$. To remove this bias, we
compute a corrected value, $\gamma$, which depends on $\gamraw$ and $\thetae$.
The \RMS resolution of $\gamma$ is about $2.5^\circ$ for
$x>0.55$ and $Q^2>5000\,$GeV$^2$.

In the quark--parton model, for a perfect detector, $\gamma$ and $E_q$
are interpreted as the
scattering angle and energy
of the massless quark $q$ in the reaction $eq\rightarrow eq$.

At a given value of $s$, the kinematic variables ($x$, $y$, and $Q^2$)
can be reconstructed from any two of the four measured quantities:
$E_e^\prime$, $\thetae$, $E_q$, and $\gamma $.  Different
combinations have been used by the HERA experiments. At
high $x$ and $Q^2$ where the calorimeter energy resolution functions are
narrow, the dominant uncertainties in energy measurements are due to
systematic effects such as energy loss in inactive material in front of the
calorimeter, nonuniformities and nonlinearities in the calorimeter response,
longitudinal energy leakages, and energy carried away by neutrinos and muons.
For the hadronic system, the raw
measured energies are typically $15\%$ less than the true energies.
For positrons, the raw measured energies are typically $4\%$ less than
the true values.

We choose the double--angle method \cite{DA} because it is least
sensitive to uncertainties in the energy measurement.  In this scheme, the
kinematic variables are obtained from $\thetae$ and $\gamma $ as follows:
\begin{eqnarray}
\xDA  &=&\frac{E_e}{E_p}\;\frac{\sin\gamma}{(1-\cos\gamma)}\;
                          \frac{\sin\thetae}{(1-\cos\thetae)}\;,
\nonumber \\
\yDA  &=&\frac{\sin \thetae(1-\cos \gamma )}
              {\sin\gamma +\sin\thetae-\sin(\gamma+\thetae)}\;,
\label{eq-davariables} \\
\QsqDA &=& s\; \xDA\; \yDA\;.
\nonumber
\end{eqnarray}

For $y>0.25$ and $x>0.45$, the resolution in $\xDA$ is $9\%$;
it improves to $6\%$ for $y>0.5$.
The resolution in $\QsqDA$ is typically $5\%$ at large $x$ and $y$.

For selected events with high $x$ and high $Q^2$
we also present the kinematic variables calculated from the
scattered positron energy $E_e^\prime$ and angle $\thetae$
using the equations:
\begin{eqnarray}
x_e &=&\frac{E_e}{E_p}\;\frac{E_e^\prime (1+\cos \thetae)}{%
2E_e-E_e^\prime (1-\cos \thetae)}\;,  \nonumber \\
y_e &=&1-\frac{E_e^\prime}{2E_e}(1-\cos \thetae)\;, \\
Q_e^2 &=&s\ x_e\ y_e\;.  \nonumber  \label{eq-evariables}
\end{eqnarray}
We apply a test--beam based correction to $E_e^\prime$ to account for
energy loss in inactive material and nonuniformities of the calorimeter
response.


\section{Event Selection}

\label{sec-evsel}

Important characteristics of reaction (\ref{eq-react}) that distinguish
it from background processes include (i) the presence
of an energetic isolated positron, (ii) $p_t$ balance, and (iii)
$E-p_Z\approx 2E_e=55\,\gev$. In addition, at large $Q^2$,
the transverse energy $E_t$ typically exceeds $100\,\gev$.

About $10^6$ events were accepted by the trigger requirements described in
section~\ref{ssec-expsetup}.
The offline event selection criteria are described below.

\begin{itemize}
\item  \uline{$E-p_Z$}\\
The net $E-p_Z$ as measured in the
calorimeter is required to be in the range $40\,\gev%
<E-p_Z<70\,\gev$ ($44\,\gev%
<E-p_Z<70\,\gev$) for $\thetae>17.2^\circ$ ($\thetae<17.2^\circ$%
). The lower cut rejects backgrounds such as photoproduction or
$e^+p\to e^+X$ events with a
hard initial state photon, for which energy escapes through the rear beam
hole (see below). The $70\,\gev$ cut removes a small number of
events with a misreconstructed vertex position.

\item  \uline{Longitudinal vertex position}\\
The event vertex
reconstructed from CTD tracks must have a $Z$ position (${Z_{{\rm vtx}}}$)
within 50 cm of the nominal interaction point. The ${Z_{{\rm vtx}}}$
distribution of the data is roughly Gaussian with $\langle {Z_{%
{\rm vtx}}}\rangle =-2\,{\rm cm}$.  The \RMS spread in ${Z_{{\rm vtx}}}$,
$12\,{\rm cm}$, is largely due to the length of the proton beam bunches.

\item  \uline{Positron requirements}\\
An isolated positron candidate with energy $E_e^\prime>20\,\gev$
and $\econe<5\,\gev$
must be found by the algorithm described in section~\ref{sec-kine}.
Additional requirements depend on the polar angle of the positron:

\begin{itemize}
\item[]  For $\thetae>17.2^\circ$, where the positron candidates are within
the CTD acceptance, a matching track with momentum above 2 GeV is required.

\item[]  For $\thetae<17.2^\circ$, where the positron either misses the CTD
altogether or is on the edge of the CTD acceptance,
the number of fake positron candidates is large. These have a
sharply falling transverse momentum spectrum.  To
reduce this background, we require positron candidates in this angular
range to have transverse momenta above $30\,\gev$.
\end{itemize}

To remove Compton scattering events ($ep\to e\gamma X$),
we reject any event which has two isolated electromagnetic clusters
in the calorimeter, each with $E_{{\rm clu}}>8\,\gev$
and $\econe<2\,\gev$.

\item  \uline{Momentum transfer}\\
We require $\QsqDA$ to exceed $5000\,\gev^2$.
\end{itemize}

The overall selection efficiency, estimated using
Monte Carlo NC events generated with $Q^2>5000\,$GeV$^2$, is $81\%$.
For the 191 events which pass
all cuts, the mean measured $E-p_Z$ is $51.9\,\gev$
with an \RMS width of $4.2\,\gev$, in good agreement
with the Monte Carlo $e^+ p$ NC simulation which predicts a mean of $51.8\,\gev$
and an \RMS of $4.0\,\gev$.
While no cut was applied to the net transverse momentum ($p_t$),
the surviving events have a mean $p_t$ of $7.5\,\gev$,
again in good agreement with the $e^+ p$ NC Monte Carlo
prediction of $7.1\,\gev$.

\section{Data and Expectations at Large $x$ and $Q^2$}

\label{sec-data}

\begin{figure}[p]
\centerline{\psfig{figure=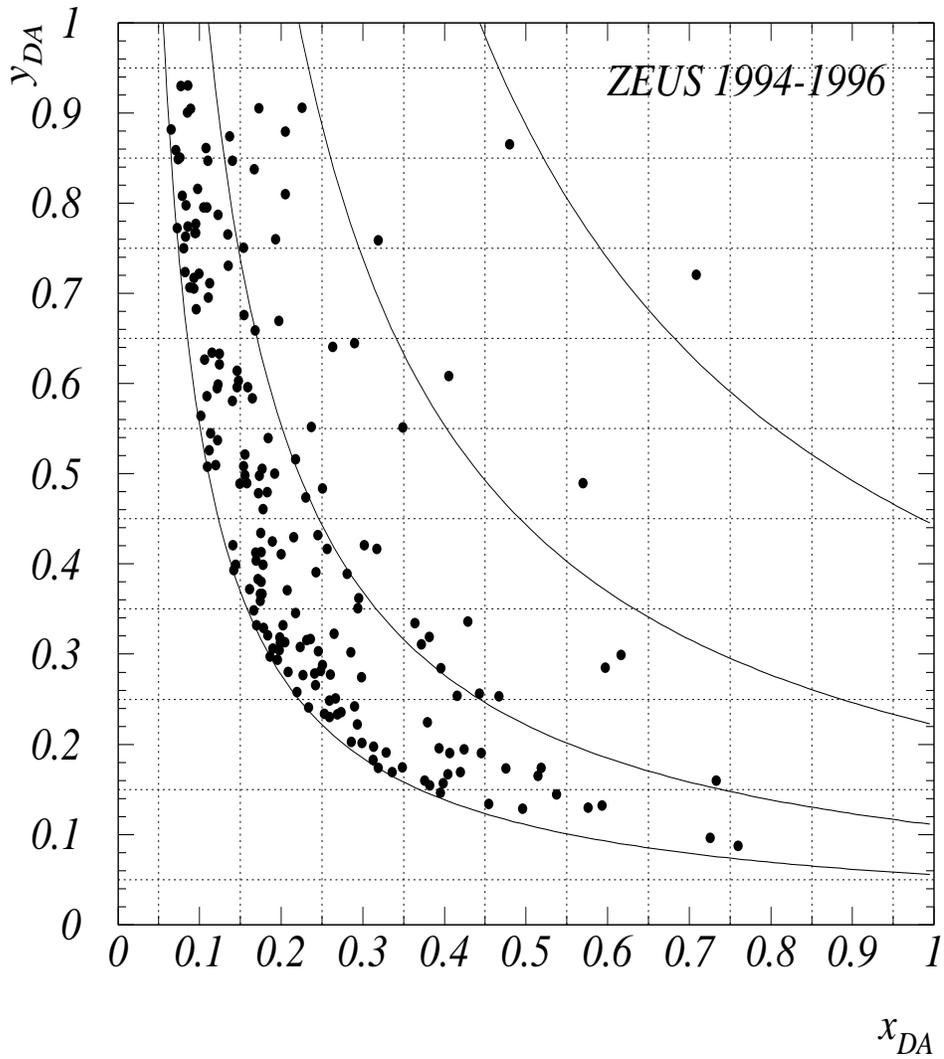,width=14cm,height=16cm}}
  \caption{
The distribution of the event sample in $\xDA$ and $\yDA$.  The
lines indicate constant values of $\QsqDA=\xDA \yDA s$ for $\QsqDA$ =
5000, 10000, 20000 and $40000\,$GeV$^2$.}
\label{fig-xyscat}
\end{figure}

Figure~\ref{fig-xyscat} shows the distribution in the $(\xDA,\yDA)$ plane
of the 191 events satisfying the selection criteria.
In Table~\ref{tab-xytab},
the numbers of observed events are compared with the Standard Model
expectations in bins of $\xDA$ and $\yDA$.
In general, the agreement between the data and the Standard Model expectations
is good. However, five events,
in four $(\xDA,\yDA)$ bins occur at high $\xDA$ and $\QsqDA$
where the expected numbers of events are small. Four lie in the region
$\xDA>0.55$ and $\yDA>0.25$, while the fifth has $\xDA=0.48$ and a
very high $\QsqDA$.
These five events are selected for more detailed discussion below.

\begin{figure}[p]
\centerline{\psfig{figure=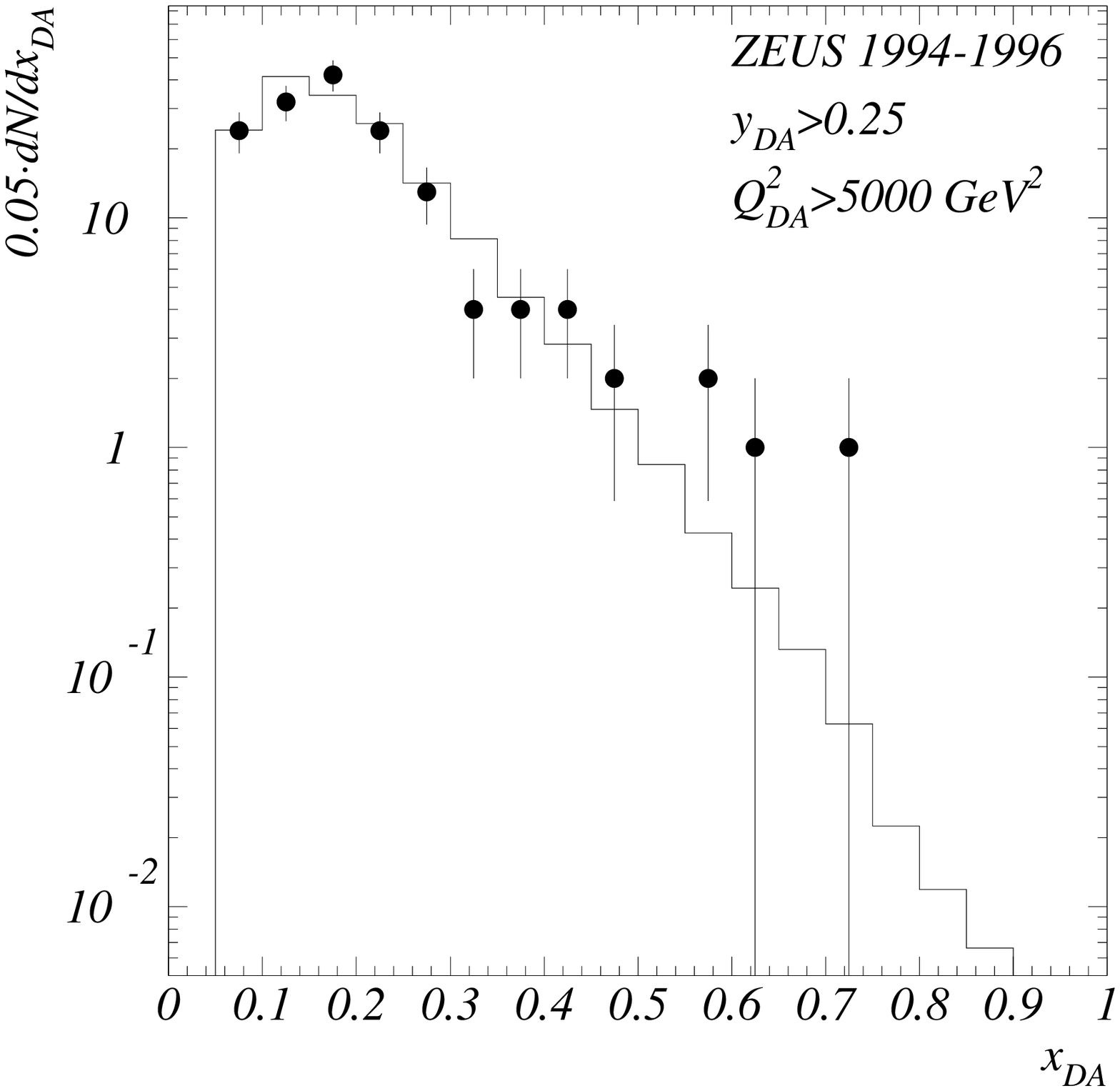,width=14cm,height=16cm}}
 \caption{
The $\xDA$ distribution of the observed events with the cuts shown (full dots),
compared to the Standard Model $e^+p$ NC expectation  (histogram).
The error bars on the data points are obtained from the square root of
the number of events in the bin.}
\label{fig-x}
\end{figure}

\begin{figure}[p]
\centerline{\psfig{figure=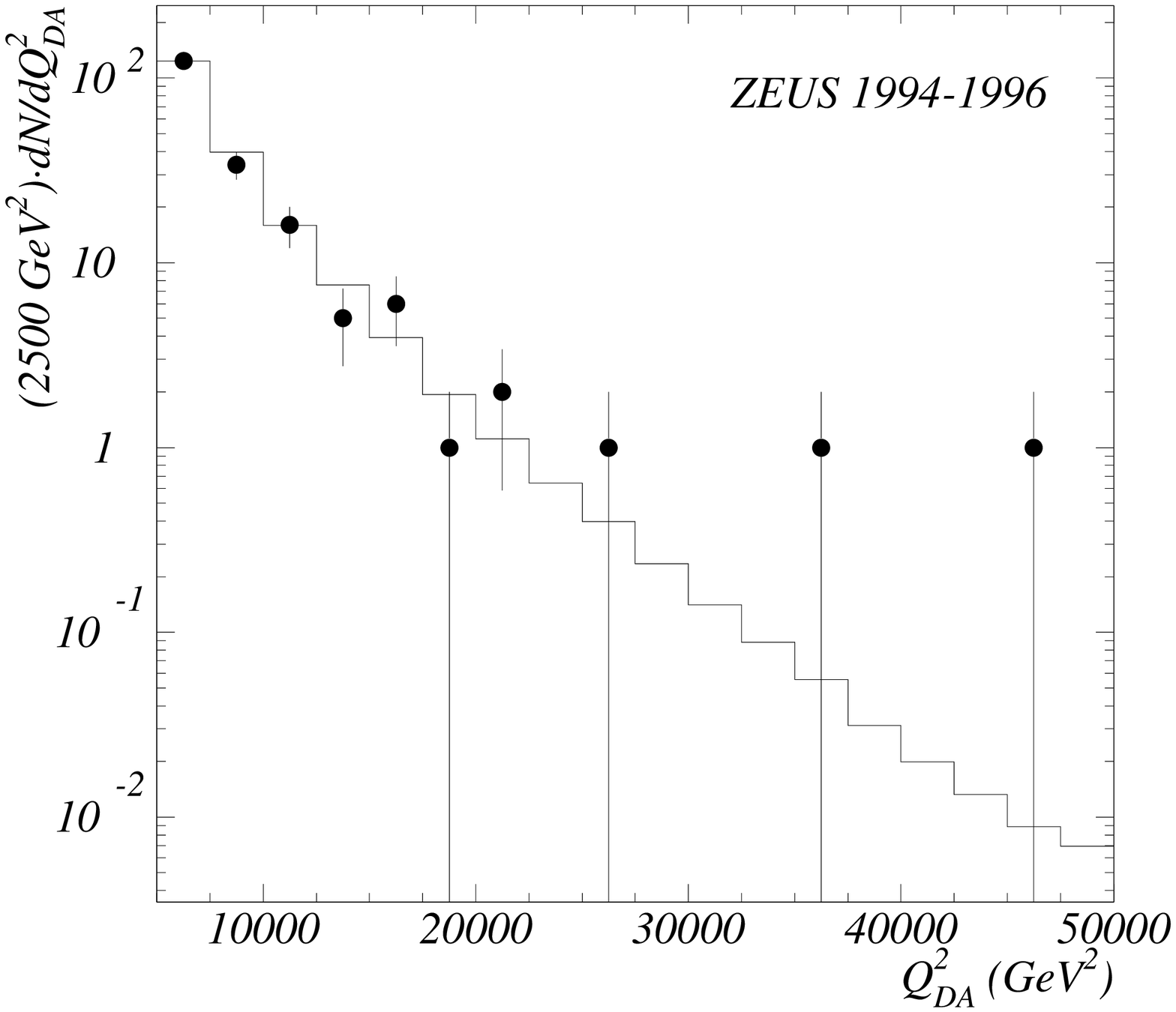,width=14cm,height=16cm}}
 \caption{
The $\QsqDA$ distribution of the observed events (full dots), compared to the
Standard Model $e^+p$ NC expectation (histogram).
The error bars on the data points are obtained from the square root of
the number of events in the bin.}
\label{fig-q2}
\end{figure}

Figures~\ref{fig-x} and~\ref{fig-q2} show the $\xDA$ (for $\yDA > 0.25$)
and $Q_\DA^2$ distributions of the final event sample.
In both figures, the $e^+ p$ NC
prediction for the same integrated luminosity is superimposed as a solid
histogram. Again, the agreement with the Standard Model is good at lower values
of $\xDA$ and $\QsqDA$, but an excess is observed at high $\xDA$ and at
high $\QsqDA$.

Table~\ref{tab-events} shows the kinematic variables,
before applying the corrections discussed in section \ref{sec-kine},
associated with the five selected
events.  Included are the uncorrected values of $\xDA$,
$\yDA$, and $Q_\DA^2$   (calculated using $\gamraw$)
as well as the corrected value of $\gamma$.
Table~\ref{tab-events_cor} gives the kinematic variables and their
estimated uncertainties  obtained using the
double--angle and electron methods.
The uncertainties have been estimated from the
resolutions in $\gamma $ and $\thetae$, as well as estimates of
the systematic uncertainty in the $\gamma$--correction procedure
discussed in Section~\ref{sec-kine}.  The
quoted \RMS errors on the electron variables
include the uncertainty in $\thetae$, the calorimeter
energy resolution, the uncertainty associated with the calorimeter
nonlinearity, and the uncertainty on
corrections applied for inactive material and nonuniformities.
Though $\thetae$ is used in both the DA and electron
methods, it makes only a small contribution to each error.
Hence the errors on the two measurements are essentially independent.

\begin{figure}[p]
\setlength{\unitlength}{1cm}
\begin{picture}(18.0,11.2)(0.0,0.0)
\put(0,0){\psfig{file=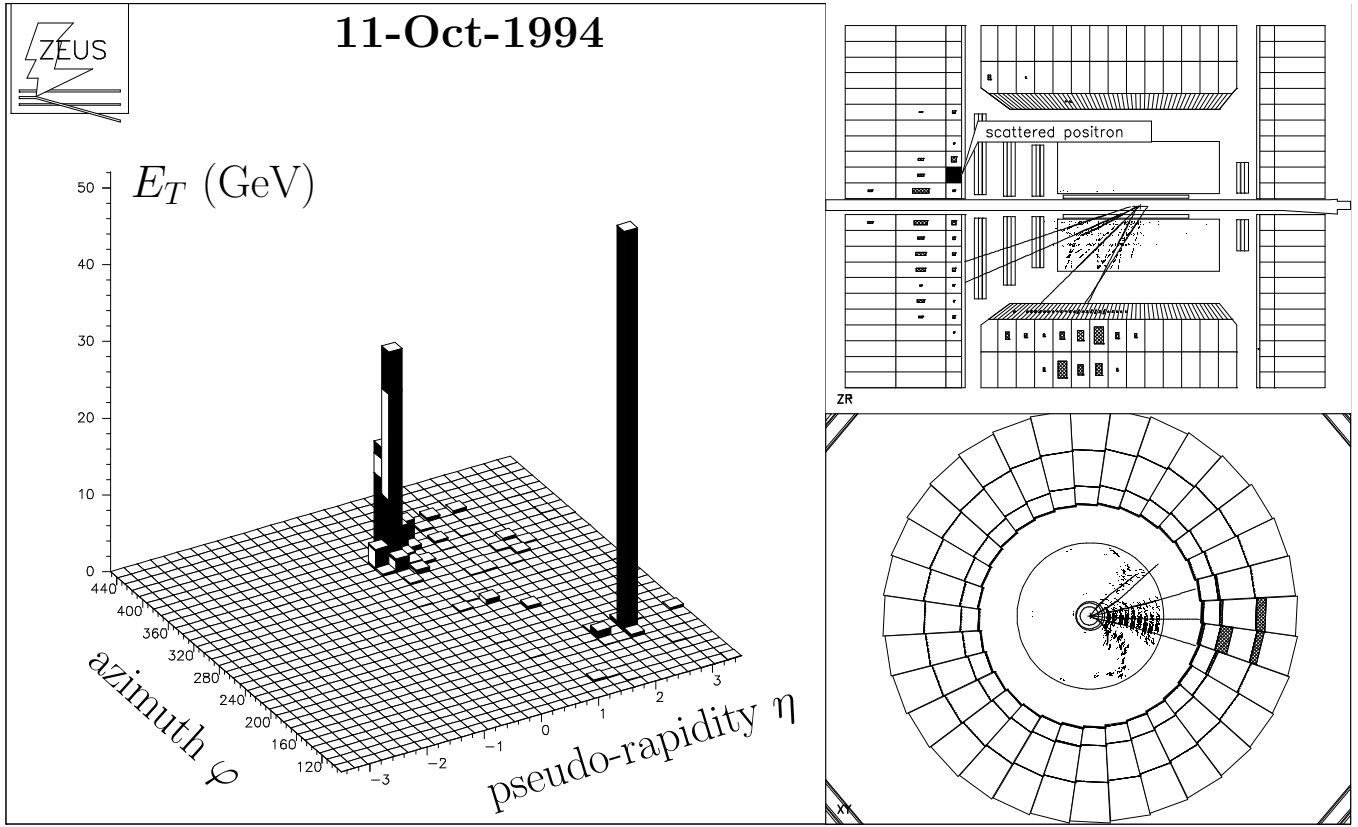,height=12cm,angle=-90}}
\put(4.0,10.5){\mbox{\Large\bf 11-Oct-1994}}
\put(1.3,8.5){\mbox{\Large $\boldmath E_T$ (GeV)}}
\put(0.7,2.3){\begin{rotate}{-42}
          \Large azimuth $\boldmath\varphi$
          \end{rotate}}
\put(6.1,0.4){\begin{rotate}{17}
            \Large pseudo-rapidity $\boldmath\eta$
          \end{rotate}}
\end{picture}
 \caption{
A display of the high $Q^2$ event recorded on 11-Oct-94. The top right
part shows the ZEUS inner tracking system and the calorimeter.
The filled rectangles in the calorimeter denote energy
deposits which are above the noise thresholds described in the text
(cf.~Section~3.1).
The bottom
right display shows a projection onto a plane perpendicular to the beam axis,
where only BCAL energy deposits are shown.  The left
part of the figure shows the calorimeter
transverse energy deposits.  This
display demonstrates that the scattered
positron is well isolated.
}
\label{fig-10012.68075}
\end{figure}

\begin{figure}[p]
\setlength{\unitlength}{1cm}
\begin{picture}(18.0,11.2)(0.0,0.0)
\put(0,0){\psfig{file=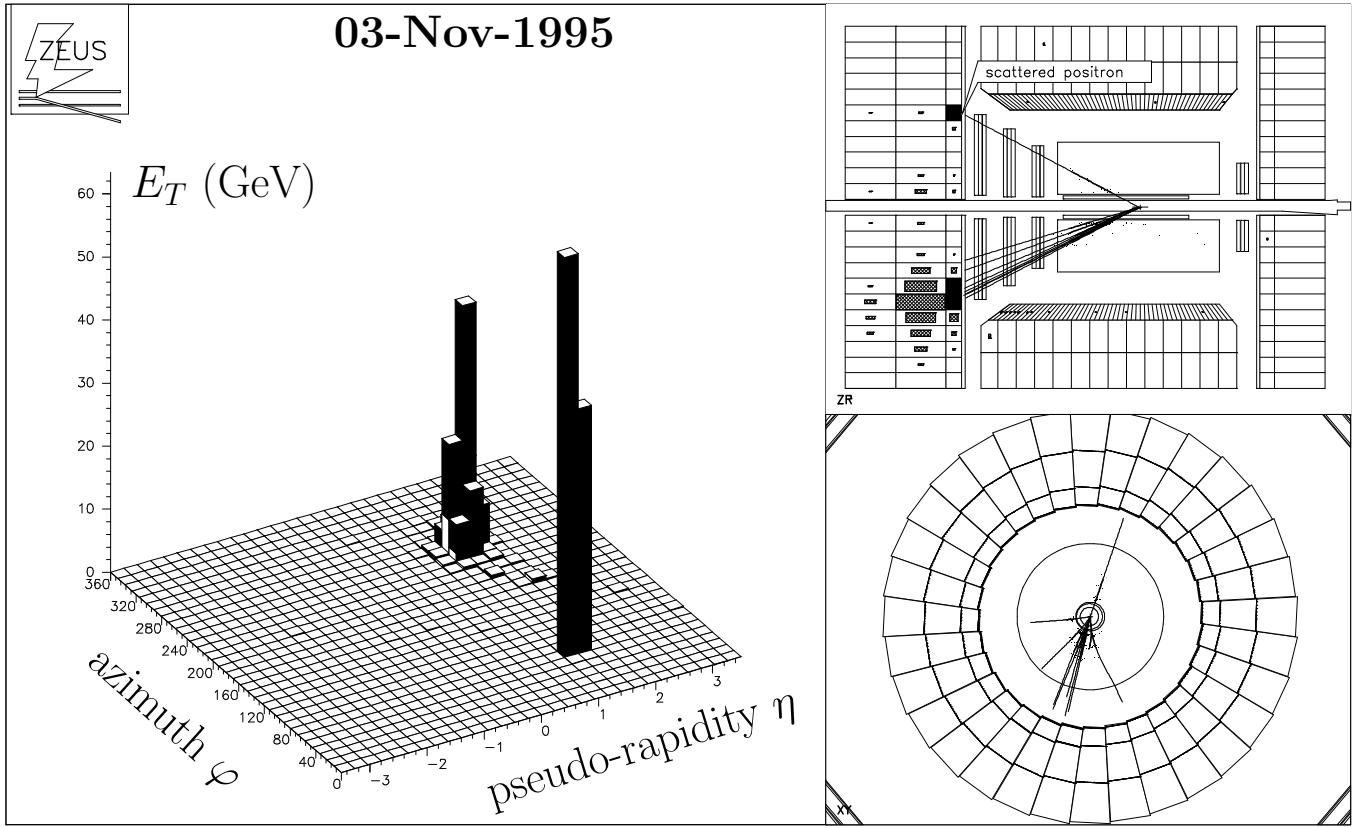,height=12cm,angle=-90}}
\put(4.0,10.5){\mbox{\Large\bf 03-Nov-1995}}
\put(1.3,8.5){\mbox{\Large $\boldmath E_T$ (GeV)}}
\put(0.7,2.3){\begin{rotate}{-42}
          \Large azimuth $\boldmath\varphi$
          \end{rotate}}
\put(6.1,0.4){\begin{rotate}{17}
            \Large pseudo-rapidity $\boldmath\eta$
          \end{rotate}}
\end{picture}
 \caption{
A display of the high $Q^2$ event recorded on 03-Nov-95. The description of
the display is identical to the previous figure. However, for this event the
positron polar angle $\thetae$ is large enough to be in the CTD
acceptance.}
\label{fig-13796.11907}
\end{figure}

All events listed in Tables~\ref{tab-events} and~\ref{tab-events_cor},
except the first, have a track matching the
electromagnetic shower of the scattered positron
in the calorimeter. In these events, the positron track momentum
is consistent with the calorimeter energy within measurement errors\footnote{
It should be noted that the positron energies in table~\ref
{tab-events} are so large that the tracking error does not allow an
unambiguous determination of the particle charge.}.
The first event (11-Oct-94) has a positron candidate at too small an
angle to produce an observable track in the CTD.\footnote{%
There are hits in the innermost layer of the CTD, aligned in azimuth with
this positron candidate. However, the hits are too few to qualify as a
track according to our standard criteria.}
We show event displays of the
first two events in Figs.~\ref{fig-10012.68075} and~\ref{fig-13796.11907}.

The five events have clean, well-identified and isolated positrons and jets
in the final state. None lie close to any of the selection cuts described in
the previous section. For these events, the scattering
angles and energies of the final state positrons and jets are measured
with good precision, making it unlikely that resolution smearing
has moved any of these events from low $Q^2$ to the measured $Q_{DA}^2$.

Initial state radiation (ISR) from the incoming positron, where the
radiated photon escapes through the rear beam hole is a possible source of
uncertainty in the determination of the event kinematics. Since ISR affects
the DA and electron variables differently, it is possible to
estimate the energy $E_\gamma$ of the radiated photon.
For each of the five events, $E_\gamma$ is consistent with zero within resolution and
the measured values of $E-p_Z$ limit $E_\gamma\lsim 3\,$GeV.

\section{Background Estimation}

\label{sec-bkgd}

Potential backgrounds to $e^{+}p$ DIS events at large $x$ and $y$ are those
processes which yield an isolated positron or electron
of high transverse energy, or a
photon or $\pi ^0$ which could be misidentified as a
scattered positron. The latter event class contributes predominantly
to the background of events in which the positron is very forward
($\thetae\lsim 17.2^\circ$) and no track information
is available for the positron candidate (e.g.\ the first event in
Tables~\ref{tab-events} and~\ref{tab-events_cor}). At larger angles,
photon conversions in inactive material between the
interaction point and the CTD can also mimic positron candidates with
matching tracks, but this effect, which is included in the detector
simulation, is much smaller.

In the following, we describe the physical processes studied
as possible sources of background.
Limits are quoted at $90\%$ confidence level.

\begin{itemize}
\item  Prompt photon photoproduction ($\gamma p\to \gamma X$) has been
studied using {\sc herwig}. We generated an event sample with the final state
photon transverse momentum exceeding $20\,\gev$. The
cross section is $1.6\,{\rm pb}$, of which
86\% (14\%) is due to direct (resolved) photoproduction.
The observed cross section due to this process
in the region $\xDA>0.45$ and $\yDA>0.25$ is 0.28 fb (0.006 events).

\item  Photoproduction of high $E_t$ jets can contribute to the background
if a jet is misidentified as a positron. Using {\sc herwig}, we
have generated event samples for both direct and resolved processes
which include heavy quark production and decay.
In these samples, no event satisfies the selection criteria for
$\xDA>0.45$ and $\yDA>0.25$,
providing an upper limit of $1.8\,{\rm fb}$ (0.04 events).

\item  QED Compton scattering ($ep\to e\gamma X$) could produce
background if one of the electromagnetic showers is not recognized
as such. Monte Carlo studies show that this probability is negligible,
with an upper limit on the contribution to the observed cross section
of $0.2\,{\rm fb}$  (0.004 events).

\item  Two photon production of lepton pairs ($\gamma\gamma\to \ell \ell$)
was studied using {\sc zlpair}.
No events from the process $\gamma \gamma \to e^{+}e^{-}$ were found
after the selections.
For $\gamma \gamma \to \tau ^{+}\tau ^{-}$, where one $\tau$ decays via
$\tau\to e\nu$, the quantity $E-p_Z$ as well as the electron transverse
energy are typically much lower than for high $Q^2$ NC events.
We obtain the upper limit on the contribution to the observed cross
section of $0.1\,{\rm fb}$ ($0.002$ events).

\item  Leptonic decays of $W$ bosons have been studied using a Monte Carlo
sample generated with {\sc epvec}. The total cross section for
production of $W^{\pm }$ bosons and their subsequent decay via $W\to e\nu _e$
is approximately $0.1\,{\rm pb}$. The final state contains a (anti)neutrino
with high transverse momentum (of order $40\,\gev$),
which typically results in large missing $E-p_Z$ (as well as $p_t$). We
estimate the accepted cross section for this process to be less than
0.5 fb (0.01 events).
Decays of the neutral boson, $Z^0\rightarrow e^{+}e^{-}$,
are rejected by the cut on two electromagnetic clusters and
are expected to contribute a negligible background.
\end{itemize}

The estimated cross sections from these background sources are
listed in Table~\ref{tab-bkgd} along with the $e^{+}p$ NC cross section. The
backgrounds are much smaller than the DIS signal in the region of
interest, and are neglected.

\section{Uncertainties of the Standard Model Predictions}

\label{sec-smerr}

The predicted numbers of $e^{+}p$ NC DIS events depend on
(i) the measured luminosity, (ii) the electroweak parameters,
(iii) electroweak radiative corrections, mainly due to initial
state radiation (ISR), (iv)  the quark densities in the relevant region
of $x$ and $Q^2$ and (v) the Monte Carlo simulation of the detector. We now
discuss the precision to which these quantities are known and describe the
studies performed to determine the uncertainties of the predictions.

\begin{itemize}
\item  \uline{Luminosity measurement}\\
The luminosity is measured to
a precision of about 1.5 \% using the ZEUS luminosity monitor.
The recent 1996 running period has a larger uncertainty
due to effects from beam satellite bunches.  Also,
the offline calibration of the
luminosity detector is not yet finalized.
Including these uncertainties from recent data, the uncertainty
for the full data sample is 2.3\%.

\item  \uline{Electroweak parameters}\\
The relevant electroweak
parameters have been measured to high accuracy \cite{EWpar} and contribute
a small uncertainty in the predicted cross section over the HERA
kinematic range \cite{ewhera}.
The {\sc heracles} program
calculates NC DIS cross sections to first order using
input values for the Fermi constant $G_\mu$, $M_Z$, the top mass $m_t$,
and the Higgs mass.
Varying $M_Z = 91.187 \pm 0.007\,$GeV and $m_t = 180 \pm 12\,$GeV
within their experimental errors \cite{EWpar}
changes the predicted cross section in
the kinematic range reported in this paper by only $0.25\%$.

\item  \uline{Radiative corrections}\\
The program {\sc hector} \cite{HECTOR}, which includes
the effects of second order QED radiative corrections
was used to check the cross sections computed using
{\sc heracles}. The differences were found to be about
$1.5\%$ for the integrated cross sections in the region $x_{DA}>0.5$ and
$y_{DA}>0.25$.

The luminosity monitor records data for all triggered events, and
so measures directly, with an acceptance of about 30\%,
the ISR spectrum for accepted events.
The experimental data are in quantitative
agreement with the ISR spectrum calculated for the accepted sample.

Corrections due to initial state radiation convoluted with the
experimental resolution, based on studies \cite{ZEUS94} made for lower
values of $x$, produce uncertainties of less than $2\%$ in
the accepted cross sections.  This number is used as the estimate of the
uncertainty due to radiative corrections.

\item  \uline{Structure functions}\\
The least well known inputs to
the predicted cross section in equation \ref{eq-csect} are
the structure functions.
To estimate the uncertainty associated with parton densities,
we performed a NLO QCD fit to fixed-target
$F_2$ lepton-proton data (with  $x > 0.1$) from
the NMC \cite{NMC}, SLAC \cite{SLAC}, and BCDMS \cite{BCDMS} collaborations
and $xF_3$ and $\bar{q}/xF_3$ results from the CCFR
collaboration \cite{CCFRqbar}.
A complete treatment of statistical and correlated experimental
systematic errors was included in the fit.
The results of the fit are consistent with the MRSA \cite{MRSA} and
CTEQ3 \cite{CTEQ} parton density parameterizations up to
$Q^2$ of $5\times10^4 \gev^2$.

The fit was used to estimate the two largest uncertainties due to the
structure functions: the experimental uncertainties and the uncertainty
of the quark-gluon coupling, $\alpha_s$, used
in the evolution to higher $Q^2$.  The effects of
experimental uncertainties in the fixed-target data
result in a $\pm 6.2\%$ uncertainty in
the integrated cross section at HERA for $x>0.5$ and $y>0.25$.
The uncertainty due to $\alpha_s$ was estimated by varying
the value of $\alpha_s(M_Z)$ used in the QCD evolution from 0.113
to 0.123, which produces an uncertainty of $\pm 1.9\%$.
From the above studies, we take
the overall uncertainty in the cross section due to
structure function uncertainties to be $\pm6.5\%$
over the kinematic range of  interest.

Other sources of uncertainty
in the structure functions were found to be small.
Changing the strange quark fraction in the QCD fit from $10\%$ to $30\%$
produced less than $0.1\%$ change in the predicted cross section. Removing
BCDMS data from the fit produced a change of only 1.7\%. Removing
data with  $W^2 = sy(1-x)$ between 10 and $25\,\gev^2$
had no significant effect.  Since the
contribution of charm to the cross section for $x>0.5$ and $y>0.25$
is $0.5\%$, uncertainties
in the charm quark mass and the charm evolution renormalization scale can
be safely neglected.

As a cross check, the uncertainty of $6.5\%$ was compared to the differences in
cross section predicted by various parton density parameterizations.
For example, a comparison of integrated cross sections
predicted by the MRSA, CTEQ3, and GRV94 \cite{GRV}
parameterizations produces an \RMS of $2\%$. A comparison of the CTEQ4
HJ parameterization \cite{CTEQhiEt} (which was tuned to the CDF high
$E_t$ jet cross section \cite{CDFanom}) with the nominal CTEQ4
parameterization produced an increase in cross section of only
$1.9\%$, demonstrating the small effect at HERA of a larger gluon
density at high $x$.  Finally, a crude estimate of the contributions
from QCD corrections at higher than NLO can be estimated by comparing
the cross sections predicted by the GRV94 LO and NLO
parameterizations, which produced a cross section difference of only
$1\%$.

Table \ref{tab-xserr_sf} summarizes the structure function uncertainties
as well as the cross checks which were performed.

\item  \uline{Detector simulation}\\
To estimate the uncertainties in
the expected event yields due to possible inaccuracies in the detector
simulation, we made several modifications to the simulation to reflect
uncertainties in the overall calorimeter energy scale and in the simulation
of the calorimeter and CTD response to positrons. The FCAL and BCAL energy
scales were separately
varied by $\pm 3\%$, our present estimate of this uncertainty.
Each of the seven measured quantities used
in the positron identification algorithm was varied by an amount consistent
with the differences between the data and the nominal simulation.
For the region $\xDA>0.55$ and $\yDA>0.25$, the resulting uncertainty in
the expected number of events is 4.4\%.

\end{itemize}

We conclude that at the large $x$ and $Q^2$ values discussed in this paper
the overall uncertainty of the number of events predicted
within the Standard Model is $8.4\%$.

\section{Comparison of Data with Standard Model and Significance of Excess}
\label{sec-signi}

Table~\ref{tab-xytab} compares the data with the $e^+p\to e^+X$
 expectations in bins of $\xDA$ and $\yDA$ for $Q^2\DA>5000\,$GeV$^2$.
There is very good agreement over the entire plane, except in the region
of high $\xDA$ and $\yDA$. The numbers of observed and expected events
above various $\QsqDA$ thresholds are given in table~\ref{tab-Q2}.
The data agree well with the Standard Model predictions up to $\QsqDA$ of
$1.5\times10^4\,$GeV$^2$.

\begin{figure}[p]
\centerline{\psfig{figure=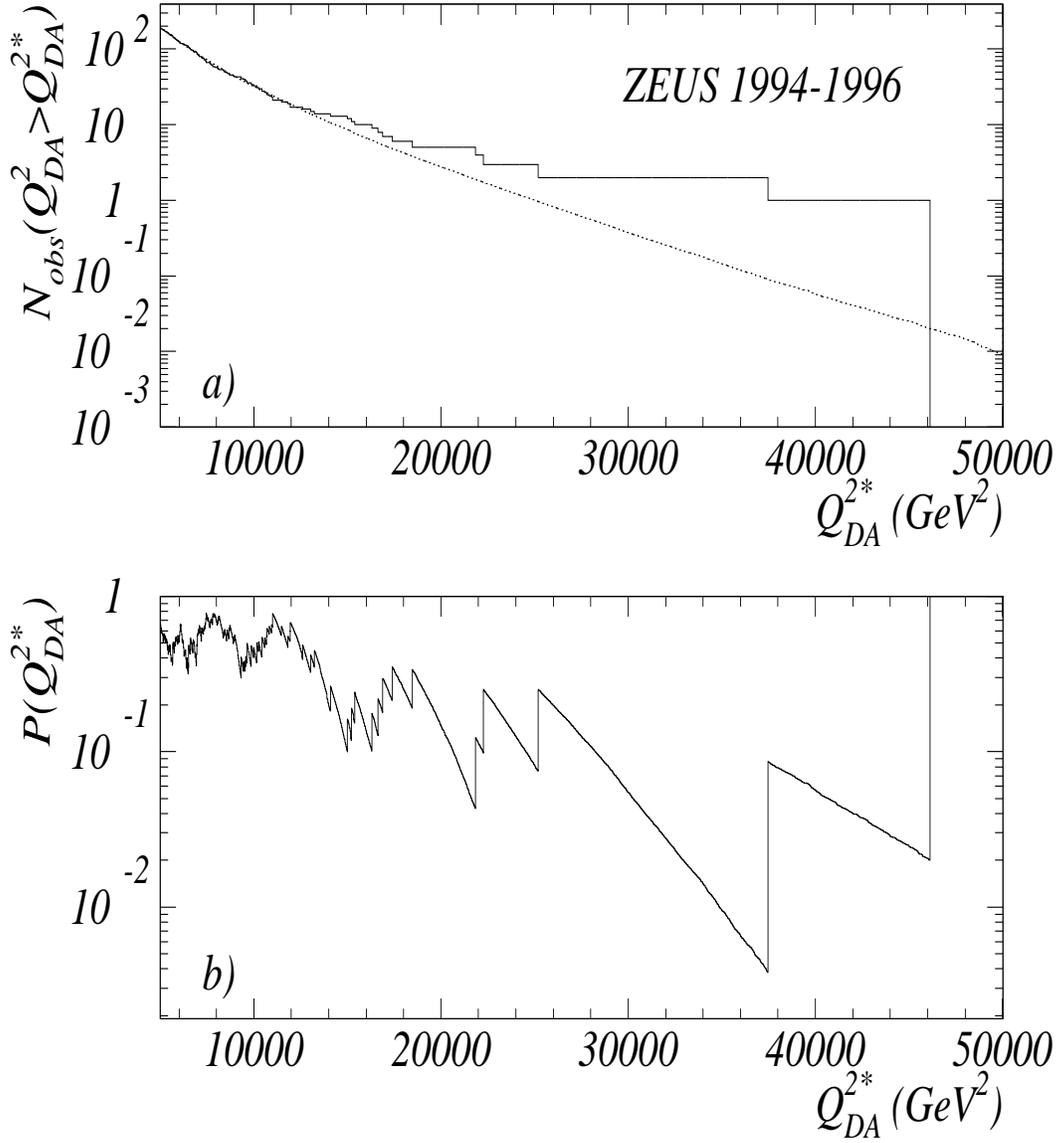,width=14cm,height=18cm}}
 \caption{
In (a), the solid line indicates the
number of observed events with $\QsqDA>\QsqDAs$ as a function of
$\QsqDAs$.
The dotted line indicates the number of events expected from $e^+p$ NC DIS
with $\QsqDA>\QsqDAs$.
In (b) is shown the Poisson probability (eqn. \ref{eq-poison})
to observe at least as many events as were observed with $\QsqDA>\QsqDAs$
as a function of $\QsqDAs$.}
\label{fig-nev_q2}
\end{figure}

\begin{figure}[p]
\centerline{\psfig{figure=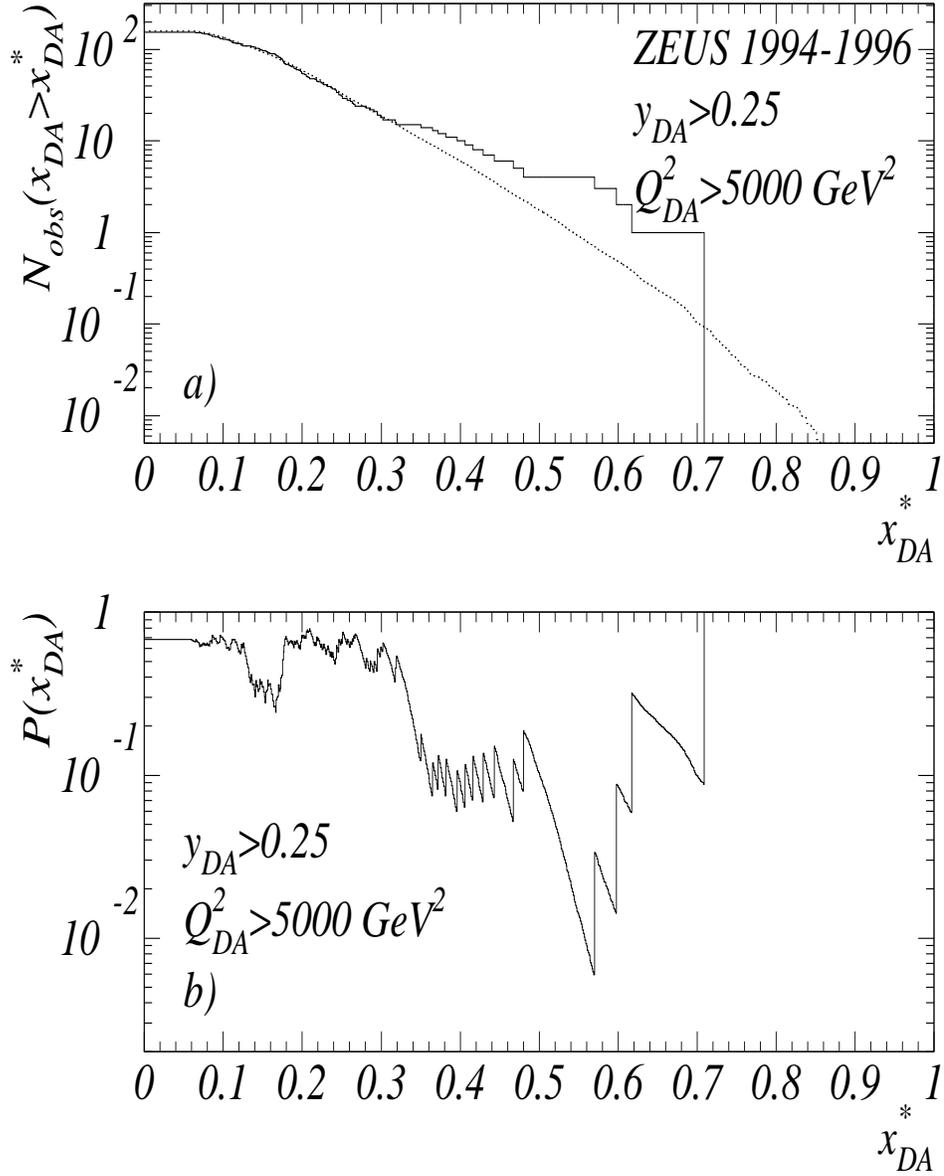,width=14cm,height=18cm}}
 \caption{
In (a), the solid line indicates the number of events observed
with $\yDA>0.25$ and $\xDA>\xDAs$ as a function of $\xDAs$.
The dotted line indicates the number of expected $e^+p$ NC DIS events
with $\yDA>0.25$ and $\xDA>\xDAs$.
In (b) is shown the Poisson probability (eqn. \ref{eq-poison})
to observe at least as
many events as were observed with $\xDA>\xDAs$ as a
function of $\xDAs$.}
\label{fig-nev_x}
\end{figure}

Fig.~\ref{fig-nev_q2}a shows the number of events with $\QsqDA>\QsqDAs$
as a function of $\QsqDAs$.
Figure~\ref{fig-nev_x}a shows the number of events
with $\yDA>0.25$ and with $\xDA>\xDAs$, as a function of $\xDAs$.
On each of the two plots, the $e^+ p$ NC DIS Monte Carlo expectation
is shown as a dotted line.

We define the Poisson probability corresponding to the event numbers
in Fig.~\ref{fig-nev_q2}a as
\begin{equation}
{\cal P}\,(\QsqDAs)=\sum_{n=N_\obs}^\infty\frac{\mu ^n}{n!}e^{-\mu}
\label{eq-poison}
\end{equation}
where $N_\obs$ is the number of observed events with $\QsqDA>\QsqDAs$, and
 $\mu$ is the number of events expected from NC DIS in the same region. In
Fig.~\ref{fig-nev_q2}b ${\cal P}(\QsqDAs)$ is shown as a function of $\QsqDAs$.
The minimum probability of
${\cal P}(\QsqDAs)=0.39\%$ (corresponding to 2.7 Gaussian standard deviations)
occurs at $\QsqDA=3.75\times10^4\,$GeV$^2$ where two events
are observed while $0.091\pm0.010$ are expected. If the expected number
of events is increased by its error, ${\cal P}(\QsqDAs)$ increases
to 0.47\%.

We have performed a similar analysis of the $\xDA$ spectrum in the
region $\yDA>0.25$. The probability ${\cal P}(\xDAs)$ is shown as a
function of $\xDAs$ in Fig.~\ref{fig-nev_x}b.
Here the minimum value ${\cal P}(\xDAs)=0.60\%$
(corresponding to 2.5 Gaussian standard deviations)
occurs at $\xDAs=0.57$ where four events are observed and $0.71\pm0.06$
are expected.  If the expected number of events is increased by its error,
${\cal P}(\xDAs)$ increases to 0.79\%.
The corresponding results for different $\yDA$ cuts appear in Table~\ref{tab-probxy}.

To gauge the significance of these probabilities, one must consider
that it is possible to observe a statistical fluctuation above {\em any }
$\QsqDAs$ or $\xDAs$ within the region studied.
We generated a large ensemble of simulated experiments according to
Standard Model assumptions, each with a luminosity of 20.1 pb$^{-1}$ and
asked how often an experiment would have a probability
${\cal P}(\QsqDAs)<0.39\%$ for {\em any} $\QsqDAs$.
The resulting probability to find such a fluctuation was $6.0\%$.
Similarly, we determined that the probability for an experiment to
have ${\cal P}\,(\xDAs)<0.60\%$ in the region $\yDA>0.25$
for {\em any} $x_{\DA}$ was 7.2\%.
The same analysis was applied for other $\yDA$ cuts and the results appear
in Table~\ref{tab-probxy}.

Finally, we have performed a statistical analysis which computes a
probability for the two--dimensional distribution of the events
in the $(\xDA,\yDA)$ plane (with $\QsqDA >5000\,$GeV$^2$).
Here the data from each simulated
experiment were binned as in Table~\ref{tab-xytab}.
Over a given region ${\cal R}$ of the $(\xDA,\yDA)$ plane,
which is defined as a subset of the bins
shown in Table~\ref{tab-xytab},
we compute the likelihood for a given experiment as
\[
{\cal L_R}=\prod_{i\in{\cal R}}e^{-\mu _i}\frac{\mu _i^{N_i}}{N_i!},
\]
where $N_i$ is the number of events observed and
    $\mu_i$ is the number of events expected in bin $i$.
For region ${\cal R}$, we denote by ${\cal L_R^{\obs}}$ the value
of ${\cal L_R}$ obtained from the data.

Using the ensemble of simulated experiments, we determined the
probability that ${\cal L_R} < {\cal L_R^{\obs}}$
for several choices of the region ${\cal R}$.
If ${\cal R}$ is the entire $(\xDA,\yDA)$ plane,
the probability that ${\cal L_R} < {\cal L_R^{\obs}}$ is $7.8\%$.
If ${\cal R}$ consists of the entire $(\xDA,\yDA)$ plane, {\em except for}
$\xDA>0.55$ and $\yDA>0.25$, the probability that
${\cal L_R} < {\cal L_R^{\obs}}$ is 50.2\%,
indicating that the data are in good agreement with the Standard
Model in this region. In contrast, the probability that
${\cal L_R} < {\cal L_R^{\obs}}$ in the
region ${\cal R}$  defined by $\xDA>0.55$ and $\yDA>0.25$  is $0.72\%$.


\section{Conclusions}

\label{sec-conc}
Using the ZEUS detector at HERA, we have studied the reaction
$e^+ p\rightarrow e^+X$ for $Q^2>5000\,\gev^2$ with a
$20.1\,{\rm pb}^{-1}$ data sample collected during the
years 1994 to 1996.

For $Q^2$ below $15000\,$GeV$^2$, the data are in good
agreement with Standard Model expectations.
For $Q^2 > 35000\,\gev^2$, two events are observed while
$0.145 \pm 0.013$ events are expected.
A statistical analysis of a large ensemble of simulated
Standard Model experiments indicates that with probability 6.0\%,
an excess at least as unlikely as that observed
would occur above {\it some} $Q^2$ cut.

For $x>0.55$ and $y>0.25$, four events are observed where
$0.91\pm 0.08$ events are expected.
A statistical analysis which  assigns a probability
to the two-dimensional distribution of the events in $x$ and $y$
yields a probability of 0.72\% for the region $x>0.55$ and $y>0.25$
and a probability of 7.8\% for the entire $Q^2>5000\,$GeV$^2$ data
sample.

The observed excess above Standard Model expectations
is particularly interesting
because it occurs in a previously unexplored kinematic region.
\vskip 2.cm \leftline{\Large\bf Acknowledgements} \vskip4.mm \noindent
We appreciate the contributions to the construction and maintenance of the
ZEUS detector by many people who are not listed as authors.
We thank the DESY computing staff for providing the data analysis environment.
The HERA machine group is especially acknowledged for
the outstanding operation of the collider.
Finally, we thank the DESY directorate for strong support and
encouragement.

\vfill\eject
\begin{table}
\begin{center}
  ZEUS 1994-1996 \\
  \vspace{0.3cm}
\begin{tabular}{|r||c|c|c|c|c|c|c|c|c|}
\hline
$x_{\rm DA}^{\rm min}$& 0.05 & 0.15 & 0.25 & 0.35 & 0.45 & 0.55 & 0.65 & 0.75 & 0.85  \\
$x_{\rm DA}^{\rm max}$& 0.15 & 0.25 & 0.35 & 0.45 & 0.55 & 0.65 & 0.75 & 0.85 & 0.95  \\
\hline\hline
$0.95<\yDA<1.00$ &  0.15   & 0.015  & 0.033  & 0.013  & 0.0055 & 0.0015 & 0.0012 &       &        \\ 
                 &         &        &        &        &        &        &        &       &        \\ 
\hline
$0.85<\yDA<0.95$ & 8.8     & 1.2    & 0.32   & 0.10   & 0.028  & 0.01   & 0.0034 &       &      \\ 
                 &  {\bf 9} &  {\bf 3}&        &        &  {\bf 1}&        &        &    &        \\ 
\hline
$0.75<\yDA<0.85$ & 12      & 2.5    & 0.50   & 0.15   & 0.050  & 0.011  & 0.0039 &       &        \\ 
                  &  {\bf 16}&  {\bf 4}&  {\bf 1}&        &        &        &        &   &        \\ 
\hline
$0.65<\yDA<0.75$ & 13      & 3.7    & 0.86   & 0.26   & 0.082  & 0.022  & 0.0054 & 0.0020 &        \\ 
                 &  {\bf 10}&  {\bf 3}&        &        &        &        &  {\bf 1}&    &        \\ 
\hline
$0.55<\yDA<0.65$ & 15      & 6.1    & 1.65   & 0.46   & 0.15   & 0.046  & 0.0090 & 0.0024 &        \\ 
                 &  {\bf 12}&  {\bf 3}&  {\bf 3}&  {\bf 1}&        &        &        &   &       \\ 
\hline
$0.45<\yDA<0.55$ & 12      &11      & 2.5    & 0.85   & 0.28   & 0.084  & 0.0208 & 0.0032 &       \\ 
                 &   {\bf 6}& {\bf 13}&  {\bf 1}&        &        &  {\bf 1}&        &   &       \\ 
\hline
$0.35<\yDA<0.45$ &  4.6    &18      & 5.5    & 1.75   & 0.52   & 0.16   & 0.0403 & 0.0093 &        \\ 
                 &   {\bf 3}& {\bf 17}&  {\bf 6}&        &        &        &        &    &        \\ 
\hline
$0.25<\yDA<0.35$ &         &18      &11      & 3.74   & 1.19   & 0.34   & 0.1104 & 0.0175 & 0.0066 \\ 
                 &         & {\bf 23}&  {\bf 6}&  {\bf 7}&  {\bf 1}&  {\bf 2}&        &  &        \\ 
\hline
$0.15<\yDA<0.25$ &         & 2.2    &14      & 9.6    & 3.32   & 1.2    & 0.2784 & 0.0717 & 0.0077\\ 
                 &         &  {\bf 1}& {\bf 15}& {\bf 10}&  {\bf 3}&        &  {\bf 1}&  &        \\ 
\hline
$0.05<\yDA<0.15$ &         &        &        & 1.3    & 2.14   & 1.6    & 0.9052 & 0.3022 & 0.1216 \\ 
                 &         &        &        &  {\bf 1}&  {\bf 3}&  {\bf 2}&  {\bf 1}&  {\bf 1}&      \\ 
\hline
\end{tabular}
\end{center}
 \caption{
The observed numbers of events in bins of $\xDA$ and $\yDA$ (bottom number
in each box), compared to the expected number of $e^+p$
NC events (top number in each box).  There are no events observed above $\xDA=0.95$.}
\label{tab-xytab}
\end{table}
\begin{table}
  \begin{center}
  ZEUS 1994-1996 \\
  \vspace{0.3cm}
  \begin{tabular}{|c||c|c|c|c|c|}
  \hline
  Event Date          &11-Oct-94&03-Nov-95&12-Sep-96&12-Oct-96&21-Nov-96\\
  \hline\hline
  $E_t\;[\gev]$       & 123.\cha& 217.\cha& 193.\cha& 204.\cha& 187.\cha\\\hline
  $p_t\;[\gev]$       &   8.9   &   8.2   &   2.9   &   2.2   &  10.2   \\\hline
  $E-p_Z\;[\gev]$     &  47.8   &  53.2   &  49.7   &  50.2   &  49.1   \\\hline
  $E_q\;[\gev]$       &  67.4   & 235.\cha& 270.\cha& 151.\cha& 276.\cha\\\hline
  $\gamraw$ &  69.0$^{\circ}$  &  28.1$^{\circ}$ & 19.9$^{\circ}$ &  40.7$^{\circ}$ &   19.7$^{\circ}$ \\
\hline\hline
  $E_e'\;[\gev]$      & 324.\cha& 220.\cha& 149.\cha& 366.\cha& 134.\cha\\\hline
  $\thetae$ & 11.9$^{\circ}$ & 27.8$^{\circ}$ & 39.3$^{\circ}$ & 15.4$^{\circ}$ & 41.1$^{\circ}$ \\ 
\hline\hline
$(\xDA)_{\rm raw}$            &   0.468 &   0.541 &   0.535 &   0.668 &   0.515 \\
$(\yDA)_{\rm raw}$            &   0.868 &   0.503 &   0.330 &   0.733 &   0.316 \\
$(\QsqDA)_{\rm raw} \; [10^4\,\gev^2]$
                      &   3.67  &   2.45  &   1.59  &   4.42  &   1.47  \\
\hline\hline
  $\gamma$ & 67.6$^{\circ}$  & 26.7$^{\circ}$ &  17.3$^{\circ}$ &  38.6$^{\circ}$ &   17.0$^{\circ}$ \\\hline
  \end{tabular}
\vskip1.cm
\caption{
Measured variables for the five events selected as described in the
text.  The first row shows the date the event was
acquired. The following
rows indicate the quantities defined in equations 7 and 8, followed by the
energy and angle of the scattered positron. The values of $x$, $y$, and $Q^2$
calculated from $\gamraw$ and $\thetae$ are shown next.
The last row shows the $\gamma$ angle.
\label{tab-events}}
  \end{center}
\end{table}
\begin{table}
  \begin{center}
  ZEUS 1994-1996 \\
  \vspace{0.3cm}
  \begin{tabular}{|c||c|c|c|c|c|}
  \hline
  Event Date          &11-Oct-94&03-Nov-95&12-Sep-96&12-Oct-96&21-Nov-96\\
  \hline\hline
  $\xDA$       &   0.480 &   0.570 &   0.617 &   0.709 &   0.597 \\
  $\delta \xDA$&   0.035 &   0.029 &   0.054 &   0.034 &   0.053 \\\hline
  $\yDA$       &   0.865 &   0.490 &   0.299 &   0.721 &   0.285 \\
  $\delta \yDA$&   0.008 &   0.010 &   0.017 &   0.008 &   0.017 \\\hline
  $\QsqDA\;[10^4\,\gev^2]$
                      &   3.75  &   2.52  &   1.66  &   4.61  &   1.54  \\
  $\delta\QsqDA\;[10^4\,\gev^2]$
                      &   0.26  &   0.07  &   0.05  &   0.16  &   0.04  \\
\hline\hline
  $x_e$       &   0.525 &   0.536 &   0.562 &   0.605 &   0.443 \\
  $\delta x_e$&   0.048 &   0.048 &   0.102 &   0.060 &   0.063 \\\hline
  $y_e$       &   0.854 &   0.505 &   0.319 &   0.752 &   0.350 \\
  $\delta y_e$&   0.018 &   0.024 &   0.039 &   0.021 &   0.032 \\\hline
  $Q^2_e\;[10^4\gev^2]$
                      &   4.05  &   2.44  &   1.62  &   4.10  &   1.40  \\
  $\delta Q^2_e\;[10^4\gev^2]$
                      &   0.31  &   0.11  &   0.09  &   0.30  &   0.07  \\\hline
  \end{tabular}
\vskip1.cm
\caption{
Kinematic variables for the five events selected as described
in the text.  The first six lines below the event dates show
the double angle values and their estimated uncertainties.  These
include the \RMS errors as well as small contributions from the uncertainties
associated with the correction procedure.
The last block of six lines shows the kinematic variables reconstructed
from the energy and the angle of the positron.  These latter errors
are dominated at present by  systematic uncertainties associated
with the positron energy measurement.
\label{tab-events_cor}}
\end{center}
\end{table}
\begin{table}
  \begin{center}
  \begin{tabular}{|c|c|c|}
  \hline
  Background&\multicolumn{2}{|c|}{cross section [fb]} \\
  Process      &$\xDA>0.45$&$\xDA>0.55$\\
  \hline\hline
  $\gamma p\to \gamma X$                       & $0.28$     & 0.28     \\
  $\gamma p\to{\rm dijets}$                    & $<1.8$    & $<1.8$  \\
  \hline
  $ep\to e\gamma X$                            & $<0.2$    & $<0.2$  \\
  \hline
  $\gamma\gamma\to\ell\ell$                    & $<0.1$    & $<0.1$  \\
  \hline
  $W\to e\nu$                                  & $<0.5$    & $<0.5$  \\
  \hline\hline
  Expected NC DIS                              & $165$     & $\cha46$\\
  \hline
  \end{tabular}
\vskip1.cm
\caption{
Expected cross sections for different background processes
in the regions $(\xDA>0.45,\yDA>0.25)$ and
$(\xDA>0.55,\yDA>0.25)$. The expected numbers of background events are
obtained by multiplying these cross sections with the integrated luminosity of
$0.02\,{\rm fb}^{-1}$.  The quoted limits are at 90\% CL. Shown for comparison
in the last row are the cross sections expected for $e^+p$ NC events.
\label{tab-bkgd}}
  \end{center}
\end{table}
\begin{table}
  \begin{center}
  \renewcommand{\arraystretch}{1.25}
  \begin{tabular}{|l|c|}
  \hline
  \multicolumn{2}{|c|}{ Systematic errors } \\ \hline
  fixed-target experimental uncertainties     & $\pm 0.062 $                \\ \hline
  $0.113 < \alpha_s < 0.123$                  & $\pm 0.019 $ \\ \hline
  \hline
  overall assumed \RMS uncertainty            & $\pm 0.065$ \\ \hline
  \hline
  \multicolumn{2}{|c|}{ Cross checks } \\ \hline
  $10\% <$ strange fraction $< 30\%$          & $ < 0.001$             \\ \hline
  uncertainties in charm evolution            & $  < 0.005$ \\ \hline
  GRV94, MRSA, CTEQ3 comparison               & $ \pm 0.020$            \\ \hline
  GRV94 NLO versus LO                         & $    +0.010$            \\ \hline
  High-$x$ gluon (CDF inspired, CTEQ4 HJ)     & $    +0.019$            \\ \hline
  \end{tabular}
\vskip1.cm
\caption{
Relative uncertainties in the integrated cross section for $x>0.5$ and $y>0.25$
due to variations in the structure functions. The top two entries represent
the two dominant contributions to these uncertainties, and so provide
the systematic
error, shown in the third row,which is used in this paper.
The remaining entries are cross checks that are not
independent of the items in the first two rows.
\label{tab-xserr_sf}}
  \end{center}
\end{table}
\begin{table}
  \begin{center}
  ZEUS 1994-1996 \\
  \vspace{0.3cm}
  \begin{tabular}{|r||c|r|r|}
  \hline
  $\QsqDAs$~[GeV$^2$] & $N_{\rm obs}(\QsqDA>\QsqDAs)$ & $\mu$ & $\delta\mu$ \\
  \hline\hline
  $5000$  & $191$ & $196.5$ & $\pm 9.87$ \\
 $10000$  & $33$ &  $32.18$  & $\pm 2.04$ \\
 $15000$  & $12$ &  $8.66$ & $\pm 0.66$ \\
 $20000$  &  $5$ &  $2.76$ & $\pm 0.23$ \\
 $25000$  &  $3$ &  $1.01$ & $\pm 0.09$ \\
 $30000$  &  $2$ &  $0.37$ & $\pm 0.04$ \\
 $35000$  &  $2$ &  $0.145$ & $\pm 0.013$ \\
\hline
  \end{tabular}
\vskip1.cm
\caption{The observed and expected numbers of events above
various $\QsqDA$ thresholds.
The first two columns give $\QsqDAs$,
the lower limit on $\QsqDA$, and $N_{\rm obs}$,
the number of observed events with $\QsqDA>\QsqDAs$.
The next two columns give $\mu$, the expected number of events
with $\QsqDA>\QsqDAs$, and $\delta\mu$, the uncertainty on $\mu$,
which includes uncertainties in the cross section prediction as
well as experimental uncertainties.}
\label{tab-Q2}
  \end{center}
\end{table}
\begin{table}
  \begin{center}
  ZEUS 1994-1996 \\
  \vspace{0.3cm}
  \begin{tabular}{|c||c|c|c|l|r|}
  \hline
  $\yDA$ range&${\cal P}_{\rm min}(\xDAs)$&$\xDAs$& $N_\obs(\xDA>\xDAs)$&$\mu$&$P_{\rm SM}$\\ 
  \hline\hline
  $\yDA>0.05$   & 1.61\%& 0.708    & 4  &  0.95\cha  & \cha16.0\%  \\\hline
  $\yDA>0.15$   & 2.57\%& 0.708    & 2  &  0.25\cha  & \cha23.0\%  \\\hline
  $\yDA>0.25$   & 0.60\%& 0.569    & 4  &  0.71\cha  &  \cha7.2\%  \\\hline
  $\yDA>0.35$   & 3.38\%& 0.708    & 1  &  0.034\cha &     26.6\%  \\\hline
  $\yDA>0.45$   & 1.32\%& 0.569    & 2  &  0.17\cha  & \cha12.7\%  \\\hline
  $\yDA>0.55$   & 0.96\%& 0.708    & 1  &  0.010\cha &  \cha9.5\%  \\\hline
  $\yDA>0.65$   & 0.50\%& 0.708    & 1  &  0.005     &  \cha5.0\%  \\\hline
  \end{tabular}
\caption{
Minimal Poisson probabilities associated with the $\xDA$ distributions for
different $\yDA$ cuts. The columns labelled ${\cal P}_{\rm min}(\xDAs)$ and
$\xDAs$ give the minimal probability and the cut on $\xDA$ where it occurs.
The next two columns give $N_{\rm obs}$ and $\mu$,
the number of events observed and the number expected with $\xDA>\xDAs$.
The column labelled $P_{\rm SM}$ gives
the probability that a simulated $e^+p$ Standard Model
experiment yields a lower value of ${\cal P}_{\rm min}(\xDAs)$
than the one observed. All values are for $\QsqDA>5000\,\gev^2$.}
\label{tab-probxy}
\end{center}
\end{table}
\vfill\eject
\end{document}